\documentclass[twocolumn,showpacs,preprintnumbers,amsmath,amssymb]{revtex4}

\usepackage{graphicx}
\usepackage{dcolumn}
\usepackage{bm}
\begin{document}
\title{Nernst response of the Landau tubes in graphite across the quantum limit}
\author{Beno\^{\i}t Fauqu\'e$^{1,2}$, Zengwei Zhu$^{1,3}$, Tim Murphy$^{4}$ and Kamran Behnia$^{1}$}
\affiliation{(1) LPEM (UPMC-CNRS), Ecole Sup\'erieure de Physique et de Chimie Industrielles, 75005 Paris, France\\
(2) H.H. Wills Physics Laboratory, University of Bristol, Tyndall Ave., Bristol, BS8 1TL, UK\\
(3) Department of Physics, Zhejiang University, Hangzhou 310027, China\\
(4) National High Magnetic Field Laboratory, FSU, Tallahassee, Florida 32306, USA}

\date {June 13, 2011}

\begin{abstract}
We report on a study of the Nernst effect in graphite extended up to 45 T. The Nernst response sharply peaks when a Landau tube is squeezed inside the thermally fuzzy Fermi surface and presents a temperature-independent fixed point when the tube flattens to a single ring. Beyond the quantum limit, the onset of the field-induced phase transition leads to a drastic drop in the Nernst response signaling the sudden vanishing of Landau tubes. The magnitude of this drop suggests the destruction of multiple Landau tubes possibly as a result of simultaneous nesting of the electron and hole pockets.
\end{abstract}
\maketitle

Manifestations of electron-electron interaction in graphene are a current subject of intense attention\cite{castro,du,bolotin}. Meanwhile, in a macroscopic stack of graphene layers, a phase transition induced by strong magnetic field was discovered three decades ago\cite{tanuma,iye1} and is believed to be driven by Coulomb interaction\cite{yaguchi1}. A large magnetic field  applied perpendicular to the conducting planes of graphite confines both electrons and holes of the semi-metal to their lowest Landau levels. A variety of instabilities can occur in a three-dimensional electron gas in such a context\cite{macdonald}. A Charge-Density-Wave (CDW) transition resulting from the 2k$_{F}$ nesting is the one which was proposed by Yoshioka and Fukuyama(YF)\cite{yoshioka} to explain the field-induced phase transition in graphite. This scenario and its variants remain the one most commonly invoked to explain this transition.

\begin{figure}
\resizebox{!}{0.3\textwidth}{\includegraphics{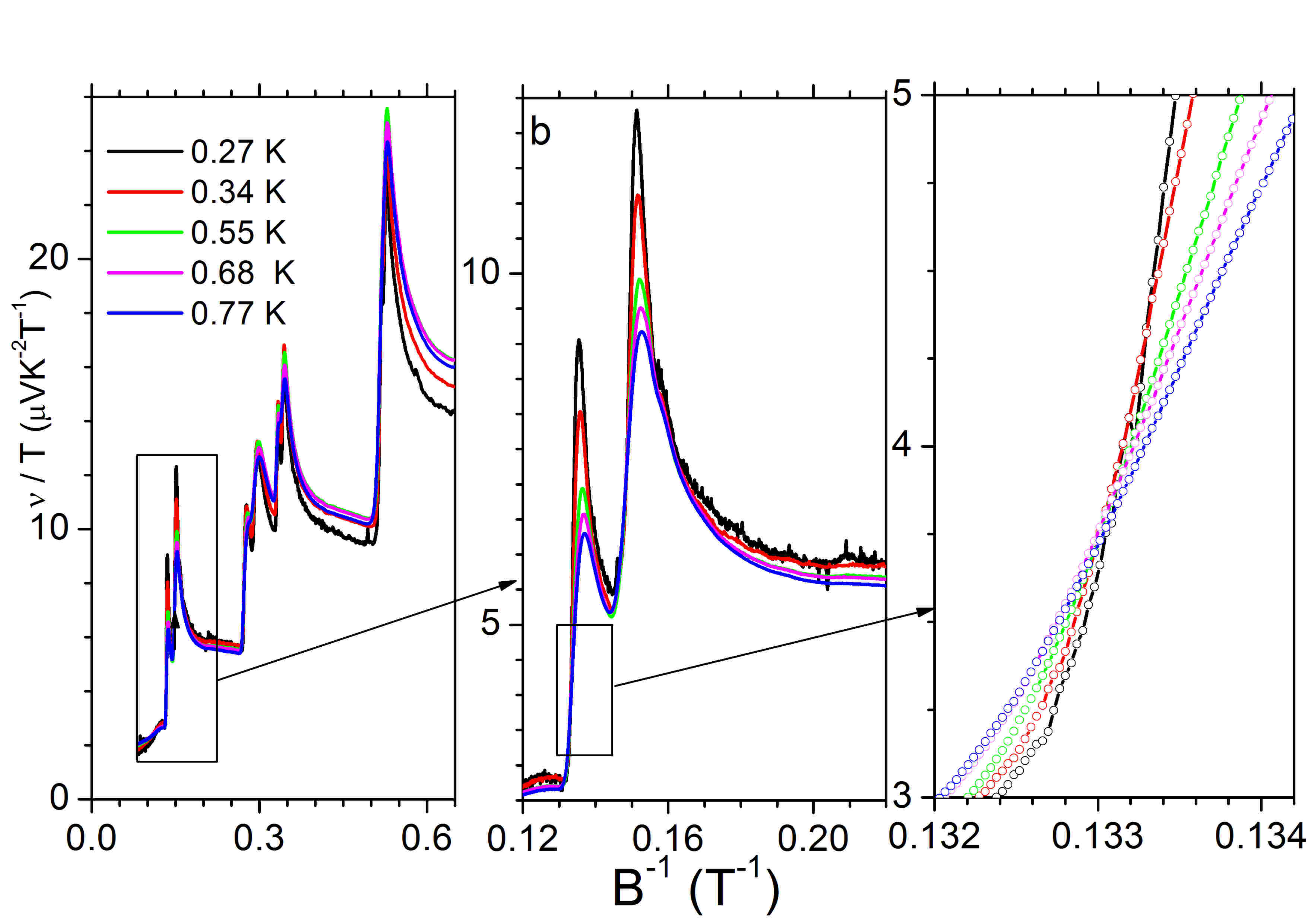}}
\resizebox{!}{0.3\textwidth}{\includegraphics{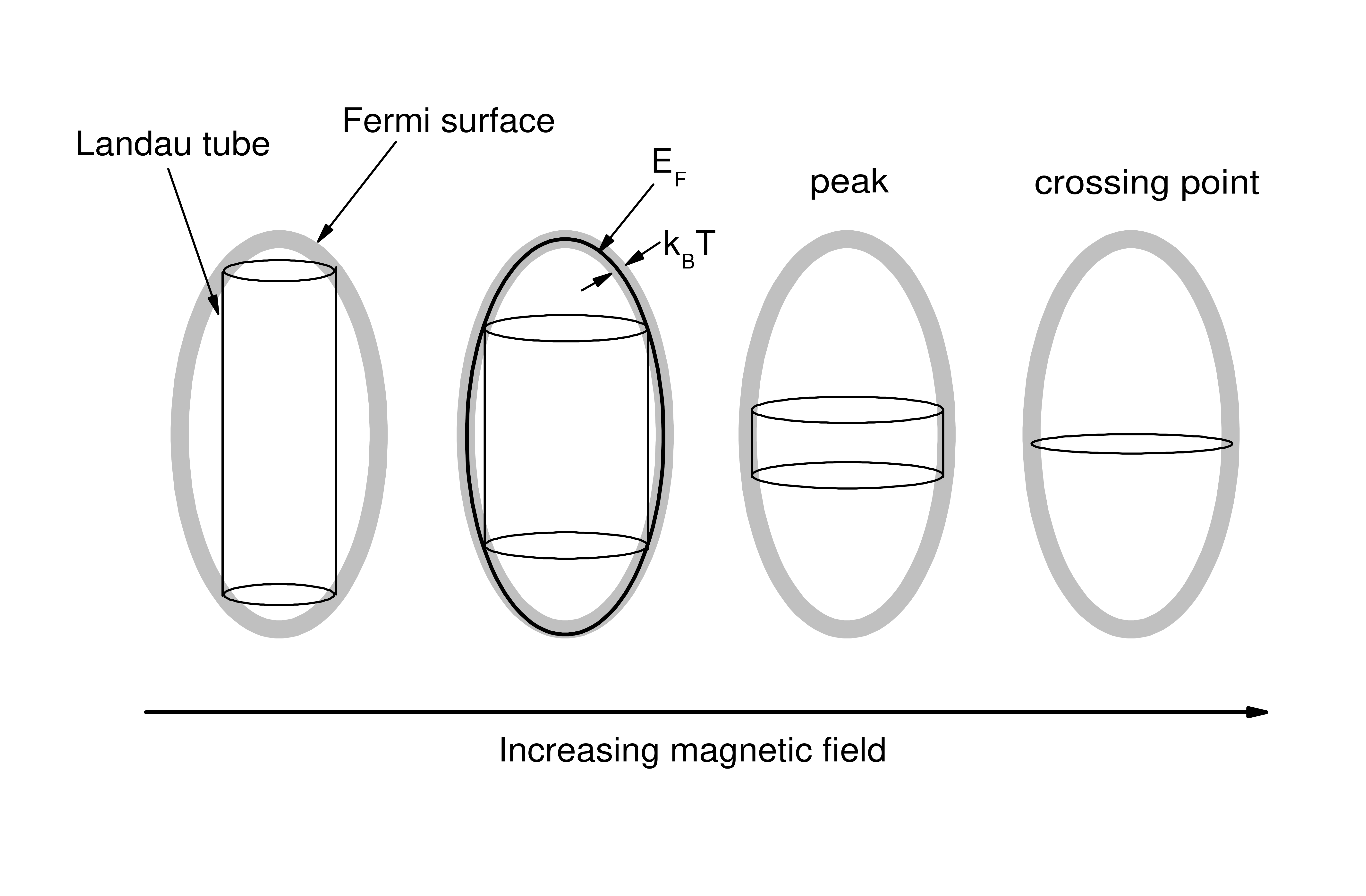}}\caption{ Top: Nernst coefficient in a HOPG sample divided by temperature in the vicinity of the quantum limit at different temperatures. Nernst peaks due to Landau levels are indexed for large (L) and small (S) components of the Fermi surface. Restricted field windows are magnified. Bottom: Each Landau tube shrinks and fattens as the field is swept.  A Landau tube squeezed inside the thermally fuzzy Fermi surface generates a peak. The field at which it flattens to a single ring is temperature-independent.}
\end{figure}

The magnitude of the field necessary to induce the transition ($\geq$ 25 T) is beyond what can be obtained by commercially available magnets. Therefore, experimental studies of this transition have been scarce and limited to resistivity measurements\cite{nakamura,iye2,ochimizu,yaguchi2,uji,yaguchi3,yaguchi4}. They have uncovered the destruction of the field-induced state at still higher fields\cite{yaguchi2}. The existence of such a reentrant transition, believed to occur when an additional Landau sub-level is depopulated, is in agreement with the YF scenario. The magnitude of the reentrant field ($\sim$ 53 T) implies significant many-body corrections to the single-particle spectrum\cite{takada}. On the other hand, a number of experimental findings do not easily fit the YF picture. Observation of nonlinear in-plane conductivity implies a CDW vector with an in-plane component, in contrast to what is expected in the simplest nesting scenario\cite{iye3}. Moreover, the structure of the phase transition appears to be complex with several transitions giving rise to different signatures in the in-plane and out-of-plane charge transport\cite{yaguchi3,yaguchi4}.

Recent experimental studies\cite{behnia1,zhu} have found that the Nernst effect is a sensitive probe of the three-dimensional electron gas in the vicinity of the quantum limit. According to a recent theoretical investigation\cite{bergman} the van Hove singularity in the density of states at the intersection of  a Landau level and a Fermi level leads to a particularly sharp peak in the transverse thermoelectric response. Fundamentally, since the Nernst response monitors the density of states per carrier, it becomes particularly  large when a large density of state is supported by a small population of electrons, which is the case of a Landau level emptying when it crosses the chemical potential.

In this paper, we report on a  Nernst study of graphite extended to 45 T, which resolves a drastic drop in the Nernst response at the onset of the high-field transition. To discuss the significance of this result, we analyze the structure of a quantum oscillation of the Nernst response as a Landau tube is squeezed and leaves the Fermi surface. We argue that the drop in the Nernst voltage at the onset of the transition can be attributed to a sudden vanishing of a pair of Landau tubes in qualitative agreement with a variant of YF scenario\cite{yoshioka,yaguchi1}. The magnitude of the drop allows us to quantify the number of the Landau tubes lost in the transition. Finally, we compare graphite with bismuth\cite{behnia2,fauque2}, which  keeps its metallicity to up to 55 T\cite{fauque2} and, instead of suffering a thermodynamic phase transition, present a cascade of additional Landau sub-levels of unidentified origin. Enumerating a number of  differences between the two systems, we discuss the possible reasons leading to the divergent fates of the three-dimensional electron gas in the two cases.

The Nernst effect was measured with a standard one-heater-two-thermometer set-up. A miniature set-up was specially designed to work in a He$^{3}$ cryostat put in the the 45 T hybrid magnet of the NHMFL. The high-field experiments were performed on two Natural Graphite (NG) samples coming from the same source as those used in a recent magnetotransport study\cite{schneider}. These measurements were complemented by detailed studies on both Highly-Oriented Pyrolitic Graphite (HOPG) and NG samples in a dilution refrigerator and a 12 T superconducting magnet. Only in the  case of low-field measurements, the absolute magnitude of the applied temperature gradient was precisely determined. In the high-field configuration, we measured the variation of the Nernst voltage produced by a constant heat flow as a function of magnetic field and checked that in the whole range of measurements phonons dominate the thermal conductivity of the samples by many orders of magnitude warranting a field-independent thermal gradient when the heat flow is kept constant.

Fig. 1 presents the profile of the Nernst quantum oscillations in graphite for the lowest Landau levels. As seen in the figure, and following previous reports on bismuth\cite{behnia1} and graphite\cite{zhu}, the Nernst coefficient  ($\nu=\frac{S_{xy}}{B}$)  presents giant quantum oscillations. In particular, there is a sharp asymmetric peak each time a Landau level empties. Let us compare the features found here with the very recent theoretical calculations by Bergman and Oganesyan\cite{bergman}.

Qualitatively, this experimentally resolved profile of the Nernst response is in agreement with theory. In particular, our data confirm their theoretical prediction of the existence of a fixed crossing point in the Nernst response. Such a fixed
point is a hallmark of quantum criticality. However, the agreement between experiment and theory\cite{bergman}, which neglects disorder and electron-electron interaction, remains qualitative. One noticeable difference is the distance between two characteristic fields; i.e. the field at which the Nernst response peaks and the field corresponding to the crossing point, where $\nu/T$ curves for different temperatures cross each other. These two field scales are distinct and significantly farther apart in our experimental data than in the theoretical curves\cite{bergman}.

\begin{figure}
\resizebox{!}{0.6\textwidth}{\includegraphics{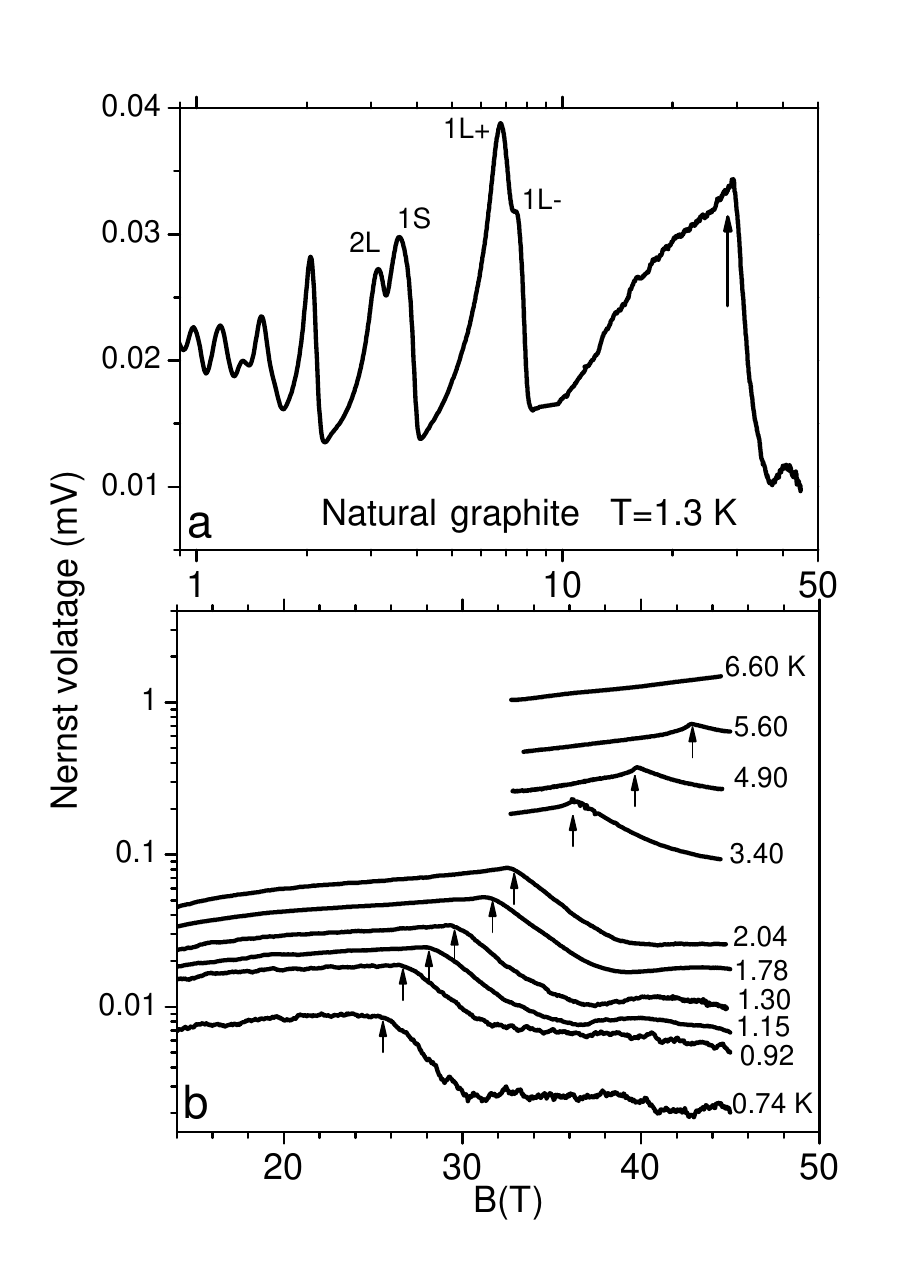}}\caption{ (a) Field dependence of the Nernst volatge in a natural graphite sample in a semi-log plot up to 45 T. Quantum oscillations below 10 T are followed by a sudden drop marked by an arrow. (b) Nernst voltage at different temperatures. Arrows track the shift in the anomaly with temperature.}
\end{figure}

Quantum oscillations arise when a Fermi surface is truncated by concentric Landau tubes \cite{shoenberg}. In the vicinity of the quantum limit, as the magnetic field is swept, the Landau tubes grow larger in diameter and shorter in length and leave the three-dimensional Fermi surface one after the other. The Nernst signal, is roughly T-linear and B-linear (i.e. $\nu/T$ remains more or less flat) when the Landau tube intersects the Fermi surface at two distinct circles. A Nernst peak occurs when the Landau tube is squeezed enough to be inside the thermally fuzzy boundary of the Fermi surface (See the bottom panel of Fig. 1). In such a configuration, the squeezed Landau tube couples an entropy reservoir with magnetic flux and becomes a source of a Nernst signal. The temperature-independent crossing point occurs when the length of the Landau tube becomes zero. After the definitive exit of the Landau tube out of the Fermi surface, $\nu/T$ drops to a value smaller than what it was before (See the upper panel of Fig. 1). The latter is a most natural consequence of the theory \cite{bergman}, as the Nernst response is a sum of the contributions by each individual Landau tube.

Fig. 2 presents our high-field data obtained on a NG sample. As seen in the upper panel, to the quantum oscillations of the Nernst voltage, succeed a drastic drop in the Nernst signal. In addition to this drop, the fine structure of the data presents smaller features, which may point to the presence of other correlation-related field scales in the extreme quantum limit\cite{kopel}. Here, we will focus on the drastic decrease marked by the arrow, which is by far the most prominent feature in our data and the undisputable thermoelectric signature of the phase transition first uncovered by resistivity measurements\cite{tanuma,iye1}. The temperature dependence of the Nernst response, presented in the lower panel of the same figure confirms this interpretation and distinguishes this anomaly from the low-field quantum oscillations. In contrast to the latter, and in a manner similar to the the jumps in the resistivity\cite{yaguchi1}, warming the sample moves the onset of the drop in the Nernst response to higher fields. We checked the presence of both these features in a second sample.
\begin{figure}
\resizebox{!}{0.4\textwidth}{\includegraphics{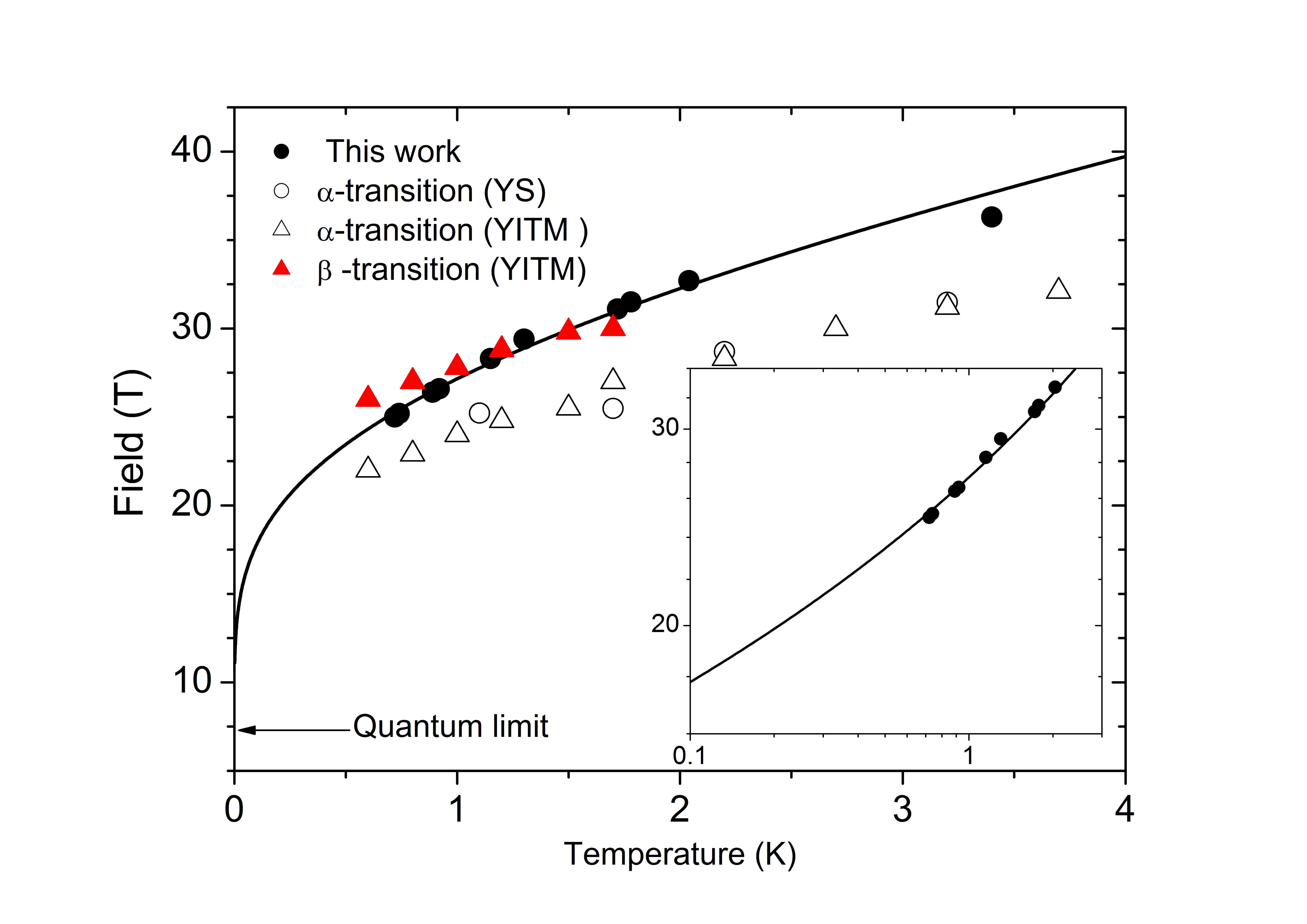}}\caption{The temperature dependence of the critical field accrording to the Nernst data (solid circles) with those obtained by resistivity measurements (\cite{yaguchi2,yaguchi3}). The solid black line is a theoretical fit to our data with the equation discussed in the text. The inset shows a semi-log plot comparing the Nernst data and the theoretical line.}
\end{figure}

Fig. 3 presents the temperature dependence of the critical field according to our Nernst data. It is comparable to what was found in various resistivity studies\cite{yaguchi1}. Its absolute magnitude is closer to the second jump in resistivity (dubbed the $\beta$ transition\cite{yaguchi2}). The temperature dependence of the transition is often described by an empirical formula inspired by the relation between a Bardeen-Cooper-Schrieffer (BSC) mean-field critical temperature, T$_{c}$ to a density of states (DOS)  proportional to magnetic field\cite{iye2}: [$T_{c}= T^{*}exp (-B^{*}/B)$]. The best fit of our data to this formula yields T$^{*}$ = 80K and  B$^{*}$= 119 T, which are comparable to what was found in previous studies\cite{iye1,nakamura,iye2}.

\begin{figure}
\resizebox{!}{0.6\textwidth}{\includegraphics{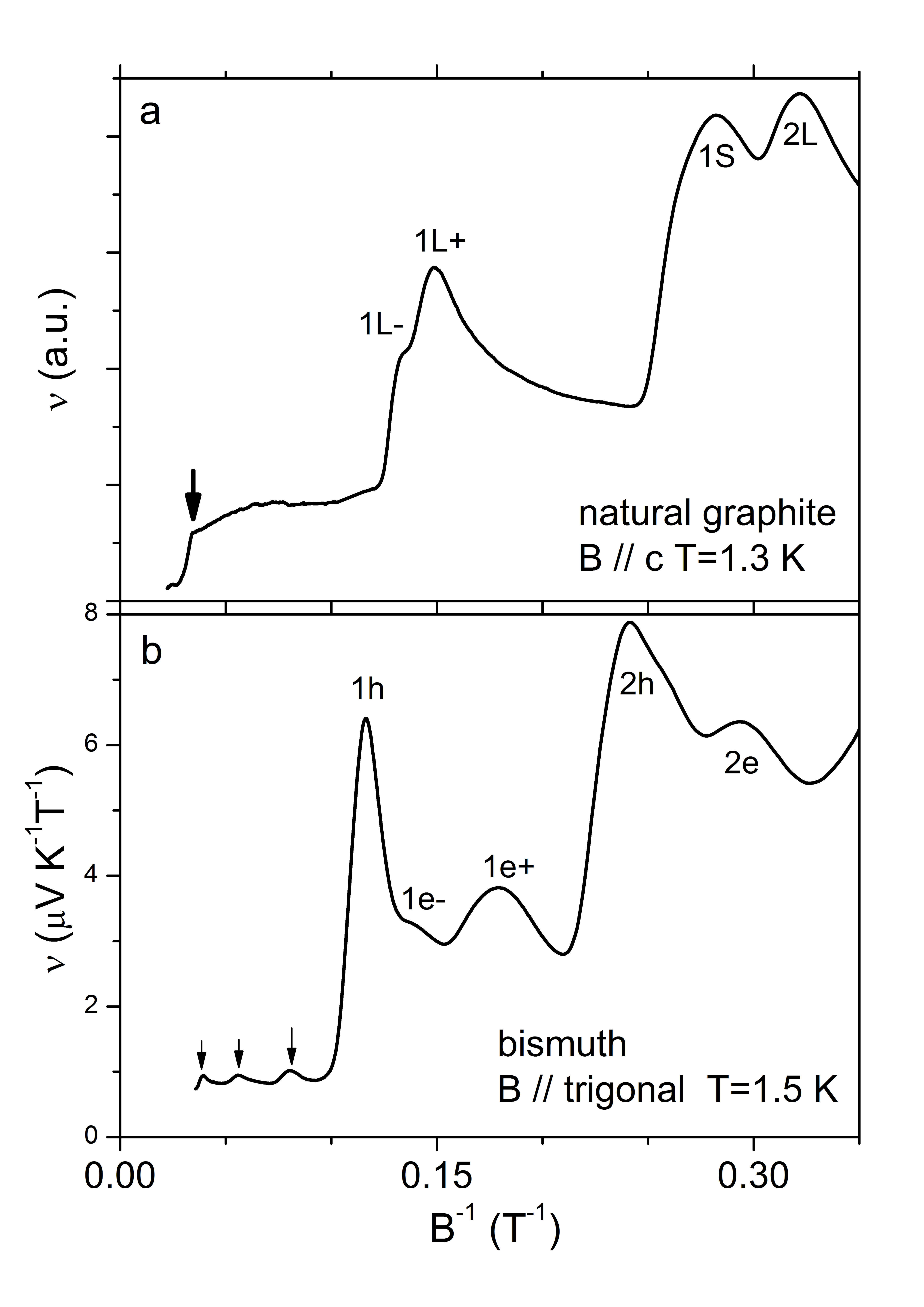}}
\caption{ (a) Field dependence of the Nernst coefficient in graphite. Nernst peaks due to Landau levels are indexed for large (L) and small (S) components of the Fermi surface. There is a sudden drop at the field-induced phase transition. (b) Same for bismuth, with peaks due to Landau levels indexed for electron (e) and hole (h) pockets. Here, there are additional peaks beyond the quantum limit, marked by arrows, which indicate unexpected additional Landau tubes.}
\end{figure}

We turn to the shape of the Nernst anomaly and its magnitude. In the low field limit, the Nernst coefficient roughly scales with the ratio of the carrier mobility to the Fermi temperature\cite{behnia3}. Naively then, one  would expect that the opening of a gap and a reduction in the carrier density should lead to a lower Fermi temperature and a larger Nernst response. In heavy-fermion URu$_2$Si$_2$, for example\cite{bel}, the unidentified phase transition which opens a gap at T = 17 K, leads to a drastic \emph{enhancement} of the Nernst response. We will argue below, however, that the experimentally-observed \emph{reduction} in the Nernst signal is \emph{compatible} with the opening of a CDW gap in the lines of the YF scenario\cite{yoshioka}.

There are two key points in our argument: i) When a Landau tube is squeezed before exiting the Fermi surface, the Nernst coefficient (that is, $\nu = S_{xy}/ B$ and not just the Nernst signal $S_{xy}$) peaks; ii)  After the definite exit of a Landau tube, the Nernst coefficient drops to a value well below what it was before the Landau tube was squeezed. Fig.4a presents the field dependence of $\nu$ according to our high-field data. As seen in the figure, the onset of the transition is associated with a sharp drop in $\nu$, in contrast to the peaks caused by quantum oscillations below the quantum limit. A natural explanation for this would be that the phase transition destroys at least one Landau tube without flattening it. This would indeed occur if the phase transition is a density-wave driven by 2k$_F$ nesting, as in the YF scenario\cite{yoshioka}.

The magnitude of the drop can be used to quantify the number of the destroyed Landau tubes. According to our data, the transition is associated with a threefold decrease in the S$_{xy}$.  According to theory, the off-diagonal Peltier conductivity $\alpha_{xy}$ is the sum of contributions by occupied Landau tubes. The contribution of each Landau tube is set by fundamental constants and the Fermi velocity\cite{bergman}. Since in graphite $\rho_{xy} \ll \rho_{xx}$, one can write $\alpha_{xy}\simeq S_{xy}/\rho_{xx}$. Thus, the threefold drop in S$_{xy}$ accompanying the thirty percent enhancement in resistivity\cite{yaguchi1} implies that $\alpha_{xy}$ drops to a quarter of its value before transition. No direct information on the effect of transition on the Fermi velocity is available, but no drastic change capable of reversing this balance is expected. Therefore, it is safe to conclude that the phase transition destroys two or three of the remaining four Landau tubes (two hole-like and two electron-like). We note that the number of Landau tubes nested in the YF scenario depends on the choice of the nesting vector, which may couple carriers with opposite or similar spins\cite{yaguchi1}.

A comparison of the Nernst response at high magnetic fields in bismuth and graphite is instructive. As seen in the panel b of the same figure, in bismuth when the field exceeds 9 T and all carriers are put in their lowest Landau level, the Nernst coefficient peaks at particular fields. At each of these fields a flattened Landau tube exits the Fermi surface. These Landau tubes are not expected in the one-particle picture and their origin is a subject of ongoing research\cite{yang}. In ultraquantum graphite, on the other hand, the Nernst data presents a drastic drop. In spite of a similar low-field departing point, there bismuth and graphite end up strikingly different at high magnetic field. Three features may cause this difference. First of all, graphite is quasi-two-dimensional. The Fermi surface consists of ellipsoids elongated across the whole Brillouin zone. Second, the mutual alignment of the two components (hole-like and electron-like) are different in the two systems. In graphite they lie along each other and in bismuth, they are almost perpendicular to each other. As a consequence, in graphite an identical nesting vector for electrons and holes is available\cite{yaguchi1}, but not in bismuth. Finally, the dielectric constant and spin-orbit coupling are much larger in bismuth.

We thank J. Alicea, D. Bergman, Y. Kopelevich, A. H. MacDonald, D. K. Maude, A. Pourret,  V. Oganesyan and H. Yang for stimulating discussions. This work is supported by the Agence Nationale de Recherche as a part of the QUANTHERM project. B.F acknowleges financial support through a Newton International Fellowship award. Research at the NHMFL is supported by the NSF, by the State of Florida and by the DOE.

\end{document}